# Catroid: A Mobile Visual Programming System for Children

Wolfgang Slany
Institute for Software Technology
Graz University of Technology
Inffeldgasse 16b, 8010 Graz, Austria

wolfgang.slany@tugraz.at

## ABSTRACT
Catroid is a free and open source visual programming language, programming environment, image manipulation program, and website. Catroid allows casual and first-time users starting from age eight to develop their own animations and games solely using their Android phones or tablets. Catroid also allows to wirelessly control external hardware such as Lego Mindstorms robots via Bluetooth, Bluetooth Arduino boards, as well as Parrot's popular and inexpensive AR.Drone quadcopters via WiFi.

## 1. INTRODUCTION
**Why programming for children?** There is a worldwide shortage of qualified software developers. This is due to rapidly increasing demand together with stagnating or even declining number of computer science students. This decline has been even more pronounced for females over the last 25 years, and it seems that even though younger girls can be interested in programming to the same degree as boys of their age, girls consistently seem to lose interest in their late teens [1,2]. At the same time our society increasingly relies on software which is thus less and less understood by the general population. Moreover, software development skills are not only of interest for obvious professional but also for philosophical reasons: Developing software is a skill that helps understanding the fundamental mechanisms and limitations underlying rational thinking.

**What is visual programming, and why do we use it?** Visual programming predominantly consists in moving graphical elements instead of typing text. We use visual programming because, based on informal experiences, it seems aesthetically to be more attractive to kids than simple text, and the success of MIT's Scratch programming environment undeniably has proven in practice more than two million times that it is very appealing to kids[1]. Note that visual programming is not easy but that if children are motivated, they are ready to spend the necessary time: Visual programming is not about dumbing down programming but instead about motivation by avoiding frustration due to, e.g., spurious syntactic mumbo jumbo, unnecessarily complicated work flows, or hard to spot syntax errors as frequently encountered in mainstream programming languages.

**A drawback of visual programming?** Visual programming has been criticized to not scale well to larger and more complex programs. However, practical evidence from visual programming environments shows that large and complex programs such as a 3D chess engine with an AI based machine opponent, multi level jump and run games, complex physics simulations, Sudoku solvers, and much more are possible with a hierarchical organization of program elements.

**Why mobile devices?** Worldwide there are ten times more mobile phones than PCs, and this ratio even is much more pronounced for children (think China and developing countries). Moreover, one's smartphone nowadays is always in one's pocket and can easily be used everywhere without preparation, e.g., when commuting to one's school using public transportation or at the backseat of the family car. Being able to program mobile devices also has become an important job qualification. Cheap smartphones from China are increasingly becoming available on a worldwide scale. Sony Ericsson's Xperia Play Android smartphone, a PlayStation based portable game console, is particularly attractive to kids.

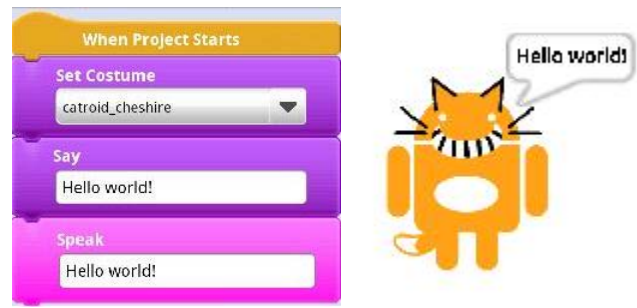

**Figure 1. Catroid's "Hello world!" program.**

---

[1] http://scratch.mit.edu/ and http://stats.scratch.mit.edu/ (2012-04)

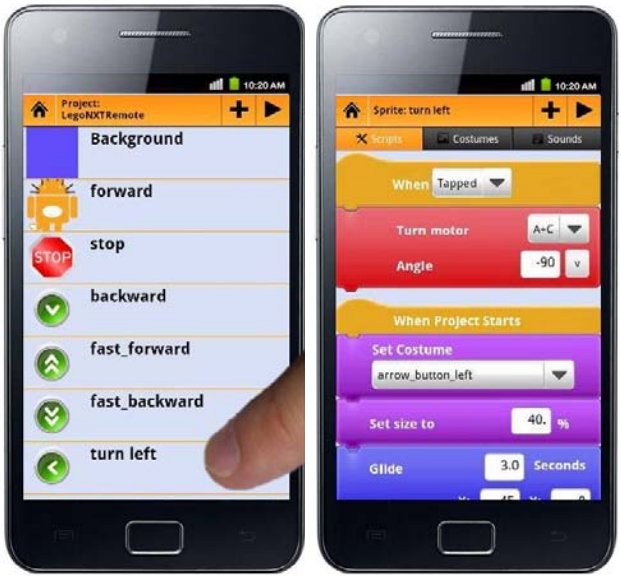

**Figure 2. Catroid program for Lego Mindstorm robots.**

## 2. SALIENT FEATURES OF CATROID

Catroid runs on smartphones and tablets, is intended for the use by children, and has been strongly inspired by the already mentioned Scratch programming language, environment, and thriving online community[1] which were developed by the Lifelong Kindergarten Group at the MIT Media Lab [4,5]. As known from Scratch or Google/MIT's App Inventor[2], Catroid programs are written in a visual Lego-style, where individual commands are stuck together by arranging them visually with one's fingers. Figure 1 on the left shows Catroid's "Hello world!" program with the Lego-style bricks sticking together. The result on the screen is shown on the right. The "Speak" brick at the bottom left of Figure 1 additionally pronounces the phrase via Android's text to speech engine in the default language of one's Android device.

Catroid also differs in important aspects from Scratch and App Inventor. Compared to both, with Catroid there is no need for a PC – the apps can be written by solely using smartphones or tablets devices. Scratch is intended for PC use with a keyboard, mouse, and comparatively large screen size whereas Catroid focuses on small devices with multi-touch sensitive screens and thereby very different user interaction and usability challenges.

As pictures often say more than a thousand words, Figures 2 to 7 show Catroid in action, thereby illustrating some of the features mentioned so far. Even better, as a session with an interactive system often says more than a thousand pictures, I cordially invite the reader to try out the latest version of Catroid[3].

Figure 2 shows parts of a Catroid program that allows controlling a Lego Mindstorm robot. On the left a list of sprite objects is shown, each possessing its own scripts and images. On the right scripts are shown that are associated with object "turn left" which is the one at the bottom on the screenshot on the left.

Figure 3 shows the resulting user interface and the robot. The necessary Bluetooth connection handshake between the robot and the Android device occurs when first executing the program. The possibilities for creative applications are infinite, especially when attaching the phone to the Lego robot and using the many sensors built into the phone such as acceleration or gyro sensors, or GPS for location based programs. Voice synthesis, voice recognition, as well as image recognition all can equally easily be used to build, e.g., autonomous intelligent soccer-playing robots. Using the similarly controlled Arduino hardware, arbitrary external devices can be controlled using Catroid.

---

[2] http://appinventor.mit.edu/

[3] http://code.google.com/p/catroid/

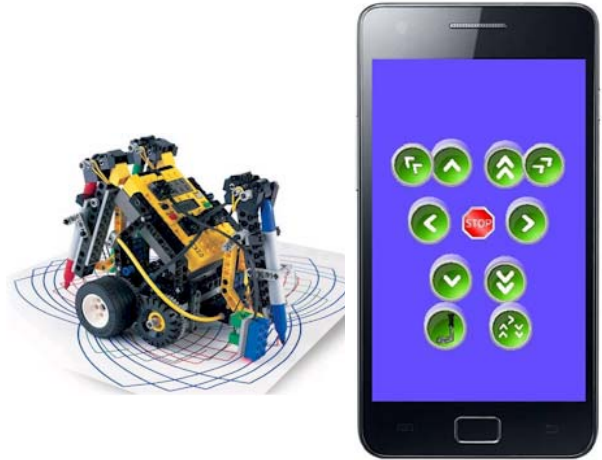

**Figure 3. Lego Mindstorm robot and user interface visually programmed with Catroid as illustrated in Figure 2.**

Figure 4 on the left shows the main screen of Catroid. On the right a part of the list of bricks that appears when one adds a brick to a script of an object is shown. The top three bricks are used for broadcasting and receiving messages, and the lower ones are used for loops and sprite movements on the screen. New command bricks selected by the user from the list of bricks partly shown in Figure 4 on the right can be visually dragged and dropped using one's fingers, or deleted by dragging them with one's fingers to a waste basket.

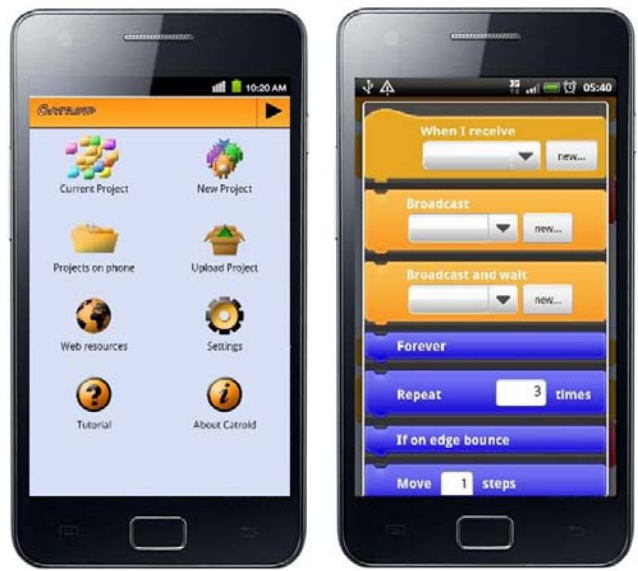

**Figure 4. Main screen of Catroid and typical command bricks.**

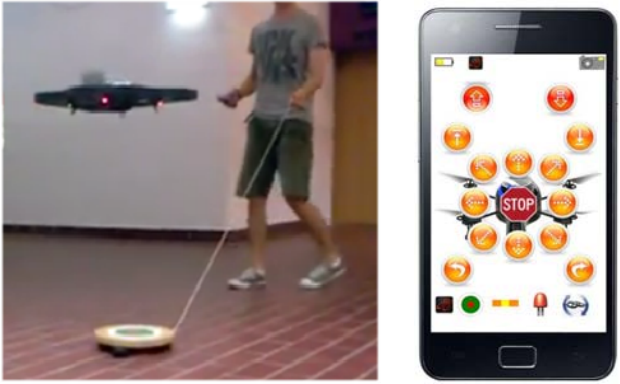

**Figure 5. Parrot's AR.Drone quadcopter controlled via WiFi by a user's program written using Catroid.**

Figure 5 shows how Parrot's popular and inexpensive AR.Drone quadcopter can be controlled from Catroid via WiFi. The quadcopter has two video cameras that transmit their data to Catroid for image processing. Catroid uses Intel's OpenCV computer vision open source library running as a service on the Android device to follow simple patterns such as the helipad in the photo on the left side of Figure 5. A video showing how it follows the moving helipad is available at http://goo.gl/1CcBK. Being able to quickly use such powerful but very simple to use features is a tremendous motivator to acquire the necessary programming skills for users of all age.

Figure 6 shows Catroid running on Sony Ericsson's Xperia Play Android smartphone, a PlayStation based portable game console. We plan to support the gamepad keys of such phones in the near future. Parents will most likely be much more willing to buy such gaming smartphones for their kids when they know that their children will not only be able to play games but moreover also be empowered to creatively build their own games, animations, simulations, or other programs.

Figure 7 shows the screen one sees when executing a Hannah Montana interactive music video animation programmed with Catroid which was created by children (it is a remix from an original Scratch project that can be found at http://scratch.mit.edu/projects/tyster/443306). Creating interactive music video animations is tremendously motivating both for boys but equally for girls even though a lot of programming is required and kids can spend days on their creations. Being able to upload such animations as videos to YouTube is an additional strong motivator as kids in general love to show off their creations to their friends regardless of what mobile phone or PC their friends are using. A YouTube recorder for Catroid programs is currently being developed, and kids will soon be able to upload their videos to YouTube in high quality. In order to allow the recording also on low-end Android devices and to decrease the amount of data that needs to be uploaded from one's Android device, only the play data is transmitted from the phone to our server. It then will be interpreted on our server in exactly the same way with the same user interaction and additional input such as random seeds. Our server records it in high quality and optionally uploads it directly to YouTube.

Similar to Scratch, Catroid is an interpreted programming language with a procedural control flow. Objects communicate via simple broadcast messages (see Figure 4 on the right) and have a set of scripts that are all excecuted concurrently, with each script running in its own thread, thus allowing real parallelism to take advantage of the multiple cores of recent smartphones. Because children easily can think in terms of objects, actors, and messages, this style of multi-threading and process synchronization feels totally natural to them. The main design objective of the language is to make it as simple to understand and use as possible.

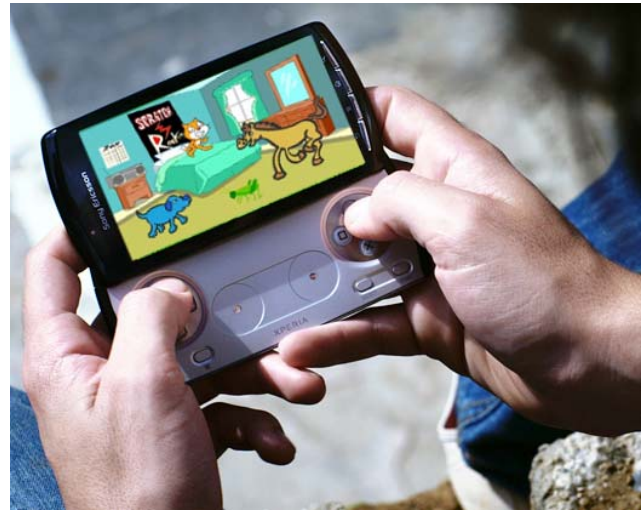

**Figure 6. Sony Ericsson's Xperia Play Android smartphone, a PlayStation based portable game console.**

The current version of Catroid as of March 2012 is not yet a full programming language as, e.g., no variables and formulas are supported at this time, though we are working hard to extend it in that direction. The project started in April 2010 with a small team. Our aim was to quickly produce a working partial solution with the most important features implemented first, and contuiune from there on. We started to implement some minimal functionality that is sufficient to emulate the highly popular creativity tool Flipnote[4] that is preinstalled on many Nintendo DSi game consoles: Only a background image that can change according to a prespecified timeline while an audio file is playing. We then went on to implement the bricks most used in Scratch programs around the world, based on statistics published on the Scratch website[5]. All of those are now implemented, and a lot more are currently being implemented to eventually make Catroid a general programming language.

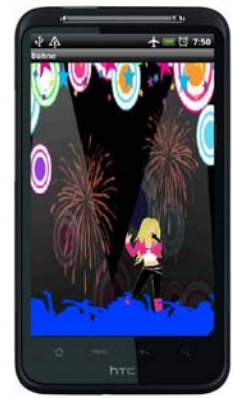

**Figure 7. Interactive music video animation.**

---

[4] http://flipnote.hatena.com/

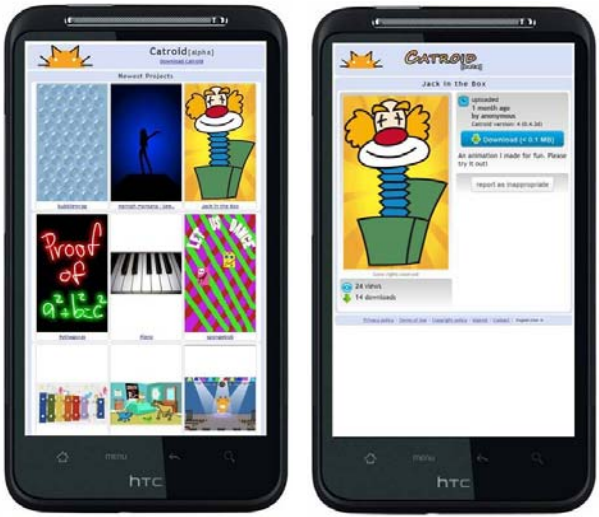

**Figure 8. Catroid's online community website.**

## 3. CATROID'S COMMUNITY WEBSITE

The Catroid system includes a community website allowing children to upload and share their projects with others. It is an important and integral part of our Catroid system. All projects uploaded to the community website are open source and published under a free software license. Everyone can download and edit every project from the website, add new functionality or change the current behavior of the project, and upload the new version again. This is called "remixing" and was a core idea behind the Scratch online community [5][5]. See Figure 8 for some images of the community website on a smartphone. On the left a list of projects on Catroid's community website is shown. On the right the details page of a project is shown. From there the project can be downloaded directly into Catroid or reported as inappropriate, as not all inappropriate content can automatically be detected. Regarding the latter, names of projects, descriptions, comments, and user names are compared to an extensive multilingual set of cuss words as well as their creative spelling variations and, when recognized, automatically rejected.

In order to serve the needs of children on a worldwide scale, both the smartphone parts as well as the website of Catroid are available in many languages. A crowd sourcing localization / internationalization support site based on Pootle allows adding further languages. We currently support several languages, with speakers of English, Mandarin, Cantonese, Hindi, Arabian, German, Turkish, French, Japanese, Urdu, Russian, Rumanian, and Malaysian in the team.

## 4. RELATED WORK

There is a plethora of research papers about visual programming: The ACM digital library lists more than 14,000 papers about the topic, and Google scholar reports more than 20,000 documents related to visual programming. I limit myself here to previous work regarding visual programming languages intended for the use by children, and in this set of languages to those featuring community sites supporting and encouraging the sharing of interactive animations and/or games created by kids. Beside Scratch and Catroid, other visual programming systems (with varying expressive power of the language) include those associated with Nintendo's Wario Ware D.I.Y.[6], Microsoft's Kodu[7] [3], Flipnote Hatena[8], and Game Maker[9] [6]. YouTube can also be seen as a platform to share user contributed multimedia content though it is not primarily oriented towards children and the contributed content cannot be made interactive.

## 5. ACKNOWLEDGMENTS

My thanks to the team members and supporters of Catroid[10].

---

[5] http://stats.scratch.mit.edu/community/blocks.html

[6] http://www.wariowarediy.com/

[7] http://www.ditii.com/2011/03/03/kodu-with-community-game-sharing-released/

[8] http://flipnote.hatena.com/

[9] http://www.yoyogames.com/

[10] http://code.google.com/p/catroid/wiki/Credits